# Generation of platicons and frequency combs in optical microresonators with normal GVD by modulated pump

VALERY E. LOBANOV[1], GRIGORY LIHACHEV[1;2], AND MICHAEL L. GORODETSKY[1;2]

[1]*Russian Quantum Center, Skolkovo 143025, Russia*
[2]*Faculty of Physics, M. V. Lomonosov Moscow State University, Moscow 119991, Russia*

**Abstract:** We demonstrate that flat-topped dissipative solitonic pulses, "platicons", and corresponding frequency combs can be excited in optical microresonators with normal group velocity dispersion using either amplitude modulation of the pump or bichromatic pump. Soft excitation may occur in particular frequency range if modulation depth is large enough and modulation frequency is close to the free spectral range of the microresonator.

Optical frequency combs generated in mode-locked lasers have revolutionized optical measurements. Last years optical microresonators have been attracting growing attention as a promising platform for microresonator-based Kerr frequency-combs [1-3]. It was demonstrated both theoretically and experimentally that frequency comb consisting of equidistant optical lines in spectral domain may appear in a nonlinear microresonator pumped by a c.w. laser due to the cascaded four-wave mixing processes. The frequency spacing between comb lines corresponds to the free spectral range (FSR, typically tens of GHz up to THz) of the microresonator which is the inverted round-trip time of light in the cavity. Microresonator-based combs were shown to perform at the level required for optical-frequency metrology applications [5] and for optical communications [6]. However, such systems often suffer from significant frequency and amplitude noise [7,8] due to the formation of sub-combs [9]. In contrast to conventional mode-locked laser-based frequency combs microresonator combs do not correspond, generally, to stable ultrashort pulses in the time domain because of arbitrary phase relations between the comb lines obtained in the process of formation. Kerr solitons solve this problem with a c.w. laser beam converted into a train of pulses, corresponding to a low-noise frequency comb having smooth spectral envelope in the spectral domain. This regime was demonstrated experimentally in optical crystalline and integrated ring microresonators with anomalous group velocity dispersion (GVD) [10, 11]. The limitation of this approach is the difficulty to obtain anomalous GVD in broad band for arbitrary centered wavelength in microresonators since material GVD in the visible and near IR is mostly normal. In this way, development of new methods that enable implementation of materials with normal GVD for frequency comb generation is interesting.

Frequency combs in normal GVD microresonators are theoretically possible [12–14]. Besides, narrow comb-like spectra from normal GVD crystalline resonators [15-17] and integrated microrings [18] have been demonstrated experimentally. Microresonators with normal GVD can also support dark optical solitons [14,19].

A novel type of solitonic pulses, "platicons", was predicted in microresonators with normal dispersion under the condition of shifted pump mode frequency resonance [20]. In real microresonators such frequency shift may occur either due to the normal mode coupling between different mode families [19, 21,22], or as a result of self-injection locking [16], both providing equivalent local anomalous dispersion [21-24]. It was demonstrated that one may continuously change in wide range the duration of generated platicons varying the pump detuning and that generation of platicons is significantly more efficient than generation of bright soliton trains in microresonators for the same absolute value of anomalous GVD in terms of conversion of the c.w. power into the power of the comb [20]. The problem with the proposed method is that it requires an artificial perturbation of the local dispersion. In this paper we show that platicon generation is possible even in the absence of the pump mode shift when bichromatic or amplitude-modulated pump is used. This method is efficient if pump modulation frequency or frequency difference between two pump waves is close to one or several FSRs. Proposed approach is experimentally feasible since comb generation from a bichromatic pump has been already studied experimentally [25] and theoretically [26] in case of anomalous dispersion. Recently, frequency comb generation by dual-pump input signal was also discussed in normally dispersive optical fibers [27].

We consider first the case of amplitude-modulated pump $|P| = P_0 \left\{ 1 + 2\varepsilon \cos\left(\Omega t\right) \right\}^2$ with modulation frequency $\Omega$ and relative amplitude of the sideband pump $\varepsilon$. Our numerical model is based on the system of dimensionless coupled nonlinear mode equations [9,28] modified to take into account the complexity of pump field. Assuming that pump modulation frequency is close to one FSR, we have:

$$
\begin{aligned}
\frac{\partial a_\mu}{\partial \tau} = &-\left(1 + i\zeta_\mu\right)a_\mu + i \sum_{\mu' \leq \mu''} \left(2 - \delta_{\mu'\mu''}\right) a_{\mu'} a_{\mu''} a^*_{\mu'+\mu''-\mu} + \delta_{0\mu} f + \\
&+ \varepsilon\left\{\delta_{1\mu} f \exp\left(i\Delta\tau\right) + \delta_{-1\mu} f \exp\left(-i\Delta\tau\right)\right\}
\end{aligned}
\tag{1}
$$

Here $\delta_{\mu'\mu''}$ is the Kronecker delta, $a_\mu$ is slowly varying amplitude of the comb modes for the mode frequency $\omega_\mu$, $\tau = \kappa t / 2$ denotes the normalized time, $\kappa = \dfrac{\omega_p}{Q}$ is the cavity decay rate, $Q$ is the loaded quality factor, $\omega_p$ is the pump frequency; $f$ stands for the dimensionless pump amplitude, $\Delta = 2(D_1 - \Omega)/\kappa$ is the normalized modulation frequency mismatch, $D_1$ is FSR. All mode numbers $\mu$ are defined relative to the pumped mode. We consider Taylor expansion of the dispersion law $\omega_\mu = \omega_0 + D_1\mu + \dfrac{1}{2}D_2\mu^2 + ...$ and neglecting third-order dispersion we get the following expressions for the normalized detunings: $\zeta_0 = 2(\omega_0 - \omega_p)/\kappa$, $\zeta_\mu = \zeta_0 + (D_2/\kappa)\mu^2$. Note, that $D_2 < 0$ for normal GVD. For numerical analysis we consider the range of parameters corresponding to experimentally feasible (realistic) microresonators: $f \in [1.5; 6.5]$ (the value $f = 1$ corresponds to bistability and comb threshold [10,14]), $D_2/\kappa \in \left[-5\times10^{-2}; -10^{-3}\right]$, $\zeta_0 \in [0;15]$. For example, for $MgF_2$ resonator [16] ($\lambda = 780\,\text{nm}$, $Q = 2.5\times10^9$, $D_2/2\pi = -2.96\,\text{kHz}$) $D_2/\kappa \approx -0.019$ and pump power of $1\,\text{mW}$ corresponds to $f \approx 3.83$; for $Si_3N_4$ microrings [18] ($\lambda = 1550\,\text{nm}$, $Q = 1.1\times10^6$, $D_2/2\pi = -300\,\text{kHz}$) $D_2/\kappa \approx -0.0017$ and pump of $1\,\text{W}$ corresponds to $f \approx 2.76$.

In our simulations we used initial weak noise-like inputs for seeding. Coupled-mode equations for 501 modes were numerically propagated in time using the Runge–Kutta integrator. Nonlinear terms were calculated using a fast method proposed in [29]. We also checked that results do not change with the increase of number of modes. For analysis we calculated average intracavity intensity $U = \sum_\mu |a_\mu|^2$ for different values of normalized detuning $\zeta_0$ [see Fig. 1(a)] and corresponding waveforms $\psi(\varphi) = \sum_\mu a_\mu \exp(i\mu\varphi)$.

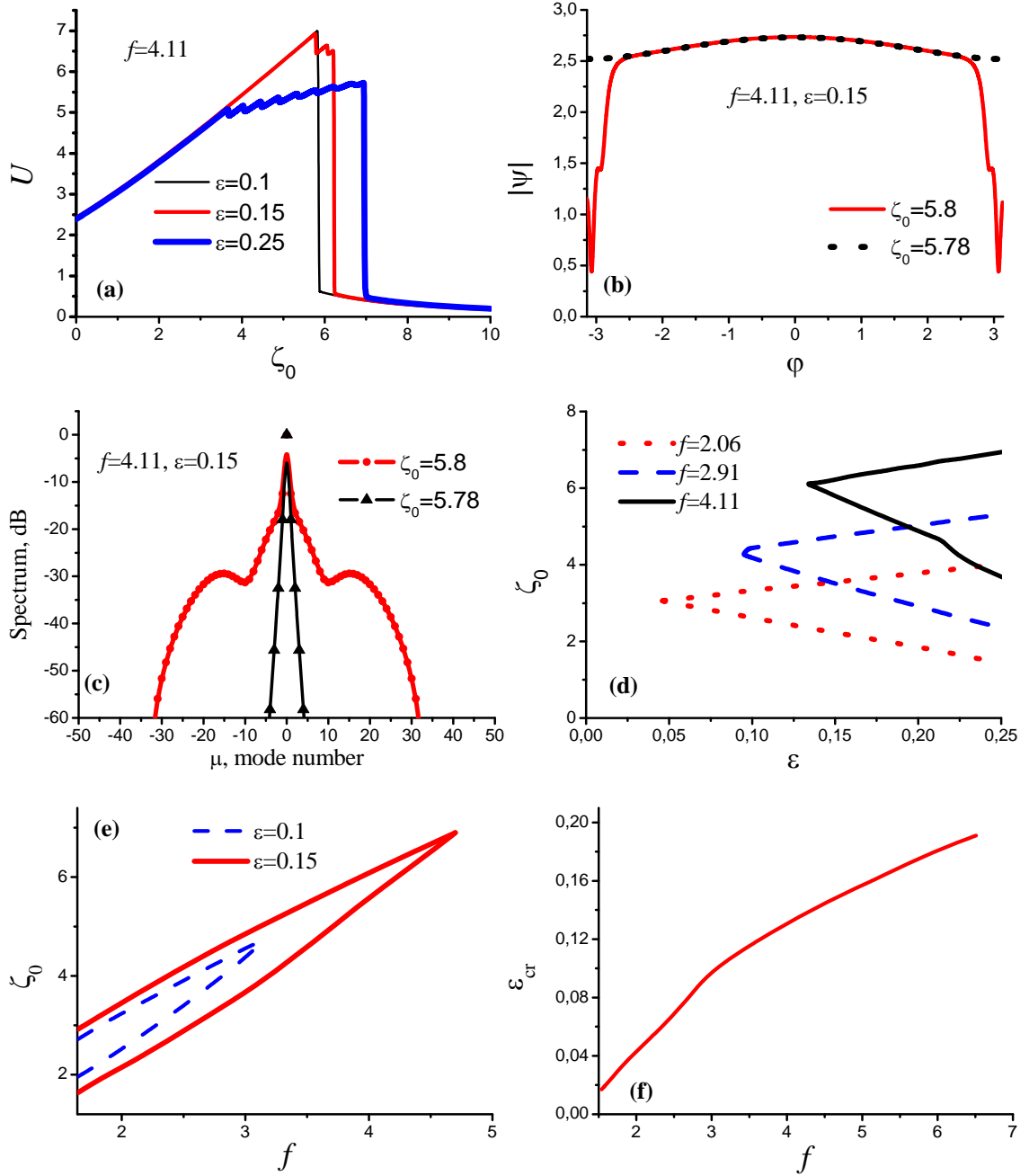

**Fig. 1.** (Color online) (a) Averaged intracavity intensity $U$ vs. normalized detuning $\zeta_0$ at $f = 4.11$ for different relative sideband pump amplitudes $\varepsilon$. (b) Profiles and (c) spectra of low-contrast wave (black line) and platicon (red line) at $\varepsilon = 0.15$, $f = 4.11$. Excitation domains (d) for different primary pump amplitudes $f$ and (e) for different relative sideband pump amplitudes $\varepsilon$. (f) Platicon generation critical value of relative sideband pump amplitude vs. primary pump amplitude. In all cases $D_2 / \kappa \approx -1.036 \times 10^{-2}$, $\Delta = 0$.

For the resonant modulation ($\Delta = 0$) it was found that at $\varepsilon \neq 0$ due to the sideband pump conventional c.w. solutions turn into low-contrast waves with narrow spectrum [see Figs. 1(b) and 1(c)]. For small sideband pump amplitudes $\varepsilon$ internal intensity vs. detuning dependence has conventional triangular shape and only low-contrast sideband modes may be observed. However, if sideband pump amplitude is large enough, this dependence deviates from the expected triangular shape, characteristic step-like dependence of intensity vs detuning appears and at particular frequency range, in the vicinity of this step, one may observe generation of high-contrast optical pulses with pronounced dip located at the minima of the low-contrast wave [see Fig 1(b)] analogous to platicons [20]. Platicon generation is accompanied by dramatic spectrum widening [Fig 1(c)]. Similarly to experimental spectra observed in [16] these are characterized by two pronounced wings [Figs. 1(c) and 2(c)]. Thus, amplitude modulation of pump power may provide soft excitation (from noise-like inputs) of platicons without tricky laser scanning necessary for bright solitons in case of anomalous GVD [10]. Such dramatic profile and spectrum transformation can be explained by the phenomenon of optical wave breaking [30] due to the double shock formation described in Ref. [27]. Shock formation is accompanied by the emergence of multiple frequency components that mix nonlinearly to produce new frequency components by four-wave mixing resulting in significand spectrum broadening.

Generated platicons were found to be stable against perturbations. To elucidate their stability we propagated them up to large periods of time (up to $\tau \sim 500$) in the presence of added white-noise-like perturbations ($U_{noise} \sim 0.1$) by solving the governing equations (Eq. (1)). Platicons rapidly (at time periods $\sim 10$) clean up the noise and propagate in a stable fashion over indefinitely large periods of time.

Platicons generated in optical microresonators look similar to "flaticons" demonstrated in optical fibers with normal GVD [31] or dispersive shock waves in an externally driven passive Kerr resonator with weak normal dispersion [32]. However, in contrast to quickly evolving flaticons and shock waves platicons are stationary soliton-like waveforms in time. It should be noted that in addition to coupled-mode approach platicons may be also described by the Lugiato-Lefever equation (driven and damped nonlinear Schrödinger equation) [14,20] as it was done for dispersive shock waves [32]. However, considered microresonators are comparatively small and pulse duration is comparable with the roundtrip time. Thus, periodic boundary conditions play an important if not determining role providing stationary solutions.

We found also that the growth of the sideband pump amplitude results in a widening of the excitation domain [Fig 1(d)]. Position of the excitation domain depends on the primary pump amplitude $f$: for larger amplitude values excitation domain shifts to larger values of $\zeta_0$ and becomes narrower [Fig 1(e)]. It is interesting to note, that for larger primary pump amplitude platicon generation requires larger values of relative sideband pump amplitude $\varepsilon$ [Fig 1(f)].

In order to study the possibility of soft excitation by phase-modulated pump we modified Eq. (1) as

$$\frac{\partial a_\mu}{\partial \tau} = -\left(1 + i\zeta_\mu\right)a_\mu + i\sum_{\mu' \leq \mu''}\left(2 - \delta_{\mu'\mu''}\right)a_{\mu'}a_{\mu''}a_{\mu'+\mu''-\mu}^* + \delta_{0\mu}f +$$
$$+\varepsilon\left\{\delta_{1\mu}f\exp\left(i\Delta\tau\right) - \delta_{-1\mu}f\exp\left(-i\Delta\tau\right)\right\}$$

(2)

and found that phase modulation does not lead to generation of platicons. Contrastingly, temporal cavity bright soliton excitation via direct phase modulation of the cavity driving field was demonstrated numerically [33] and experimentally [34].

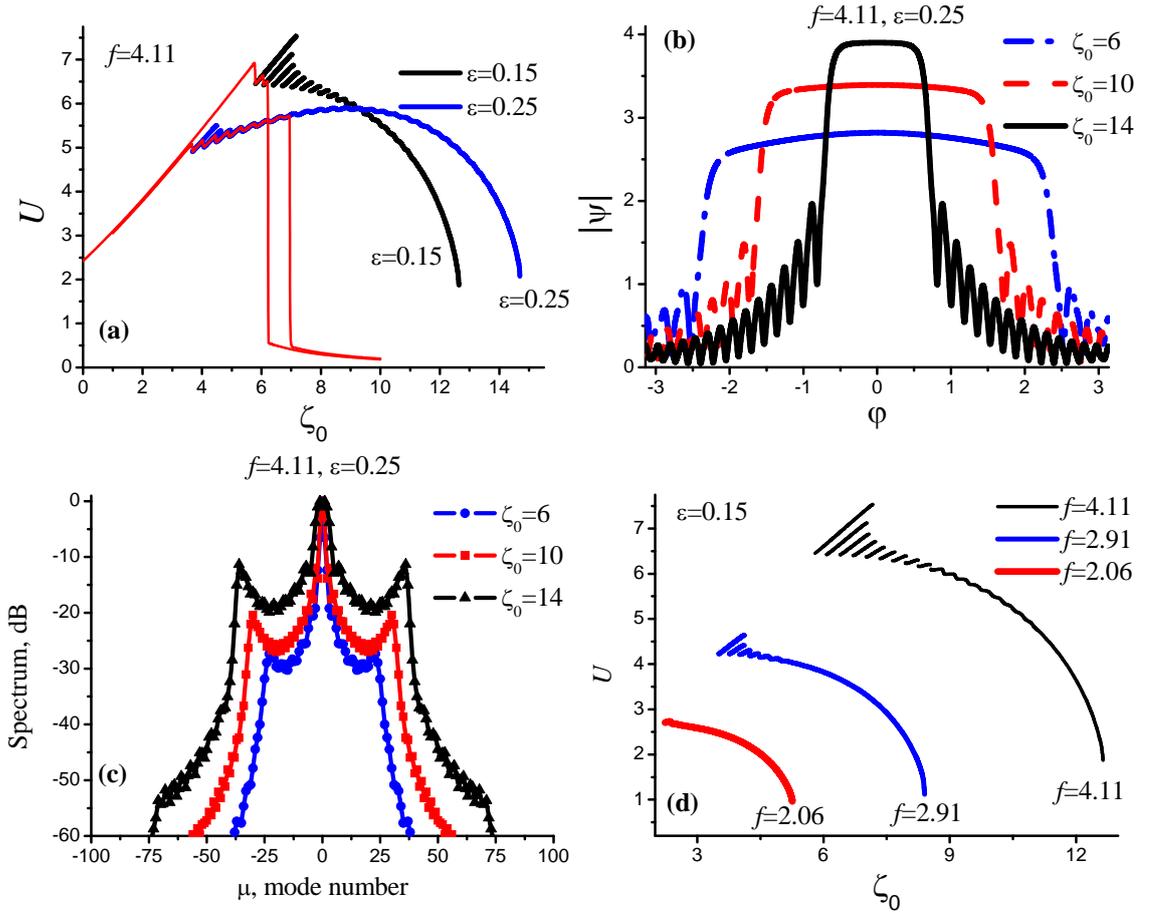

**Fig. 2.** (Color online) (a) Averaged intracavity intensity $U$ of platicons vs. normalized detuning $\zeta_0$ at $f = 4.11$ for $\varepsilon = 0.15$ (black line) and $\varepsilon = 0.25$ (blue line). Thin red lines indicate the results of soft excitation. (b) Profiles and (c) spectra of platicon for different values of $\zeta_0$ at $\varepsilon = 0.25$, $f = 4.11$. (d) Averaged intracavity intensity $U$ of platicons vs. normalized detuning $\zeta_0$ for different values of $f$ at $\varepsilon = 0.25$. In all cases $D_2 / \kappa \approx -1.036 \times 10^{-2}$, $\Delta = 0$.

Using an initial platicon as an input for our simulations we found all possible platicon solutions [Fig. 2(a)]. Interestingly, the excitation domain, defined as detuning range where platicon excitation from noise-like inputs is possible, is significantly narrower than the full existence domain and lies at its lower bound (in terms of $\zeta_0$) [compare widths of characteristic steps at thin red lines at Fig. 2(a) corresponding to excitation domains and widths of existence domains where thick blue and black lines lie]. Moreover, it was revealed that noise-like inputs provides excitation of wide platicons only. However, having wide platicon, generated inside the excitation domain, one may further find all other possible platicon solutions varying slowly pump frequency (or detuning value $\zeta_0$) and, consequently, may determine full existence domain. It should be noticed that if several platicon solutions may exist at the same pump frequency, soft excitation provides generation of the low-power platicon.

It was found that while the sideband pump amplitude increases, the existence domain of platicons becomes wider and energy spectrum contains smaller number of steps. In this way, in considered system parameter $\varepsilon$ behaves analogously to the parameter describing pump mode shift in [20] and similarly growth of $\varepsilon$ results in transformation of discrete energy spectrum of dark solitons into quasi-continuous spectrum of platicons.

We revealed that similarly to platicons described in [20] one can effectively control the duration and, consequently, the power of generated pulses by slowly tuning the pump frequency. Figure 2(b) shows that while pump frequency decreases (corresponding detuning $\zeta_0$ increases), the pulse duration also decreases. Consequently, the width of the corresponding frequency comb may also be tuned [Fig. 2(c)].

It was found that while the pump power decreases, the existence domain of platicons becomes narrower and shifts to smaller values of $\zeta_0$ [Fig. 2(d)]. In this way, one can tune platicon duration varying primary pump power: decreasing pump one may generate shorter pulses.

To prove that proposed approach is applicable for a wide range of materials used for fabrication of microresonators with normal GVD we make calculations for a wide range of GVD values from $D_2 / \kappa \approx -1.036 \times 10^{-3}$ up to $D_2 / \kappa \approx -4.142 \times 10^{-2}$. It was found that growth of absolute value of $D_2$ results in weak increase of sideband pump power necessary for platicon generation [see Fig. 3(a)]. Decrease of the absolute value of the dispersion coefficient results in more localized pulses with sharper profiles and wider spectrum [Figs. 3(b) and 3(c)]. The positions of spectrum peaks and, consequently, spectrum width may be estimated rather accurately by the simple phase-matching condition [35]: $\zeta_\mu = \zeta_0 + (D_2 / \kappa)\mu^2 = 0$.

Also, while existence domain weakly depends on $D_2$, increase of absolute GVD value leads to more pronounced discreteness of platicon energy spectrum [Fig. 3(d)].

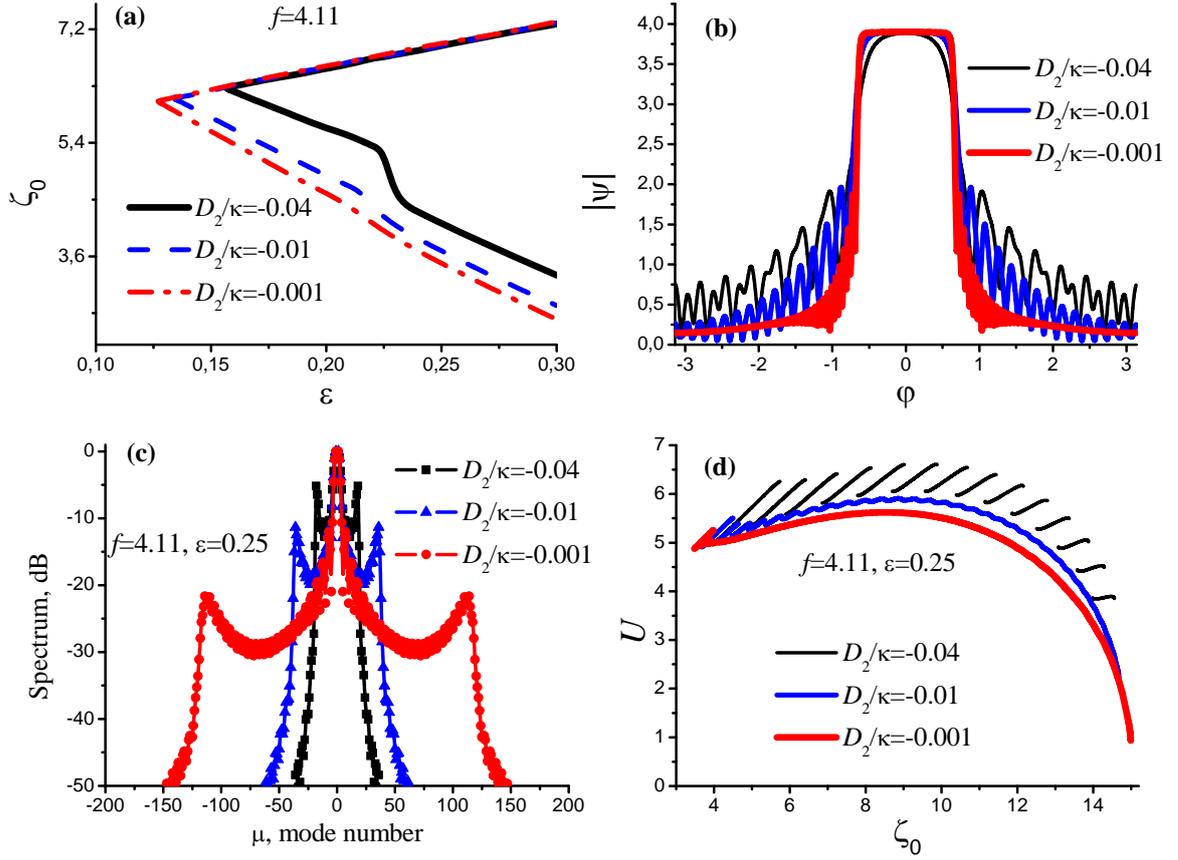

**Fig. 3.** (Color online) (a) Excitation domains for different values of GVD at $f = 4.11$. (b) Profiles and (c) spectra of platicon for different values of GVD at $\varepsilon = 0.25$, $f = 4.11$, $\zeta_0 = 14$. (d) Averaged intracavity intensity $U$ of platicons vs. normalized detuning $\zeta_0$ at $f = 4.11$, $\varepsilon = 0.25$ for different values of GVD. In all cases $\Delta = 0$.

We also found that the method considered is very sensitive to the pump modulation frequency. While the mismatch between modulation frequency and FSR increases, existence [Fig. 4(a)] and excitation [Fig. 4(b)] domains become significantly narrower and disappear at some critical value of mismatch. The value of mismatch critical for platicon excitation increases with sideband pump amplitude $\varepsilon$ [see Fig. 4(b)]. This may be explained by the inefficiency of sideband pumping at large mismatches. In this came pump becomes effectively monochromatic and, thus, can not provide soft excitation of platicons or dark solitons at considered dispersion law as it was shown earlier [14,20].

It was also revealed that for frequency combs generated by amplitude-modulated pump the spacing between the comb lines is equal to the modulation frequency but not to the FSR of a microresonator. Interestingly, but not surprisingly, if the modulation frequency is equal to an integer number of FSR, the corresponding number of platicons per roundtrip can be generated.

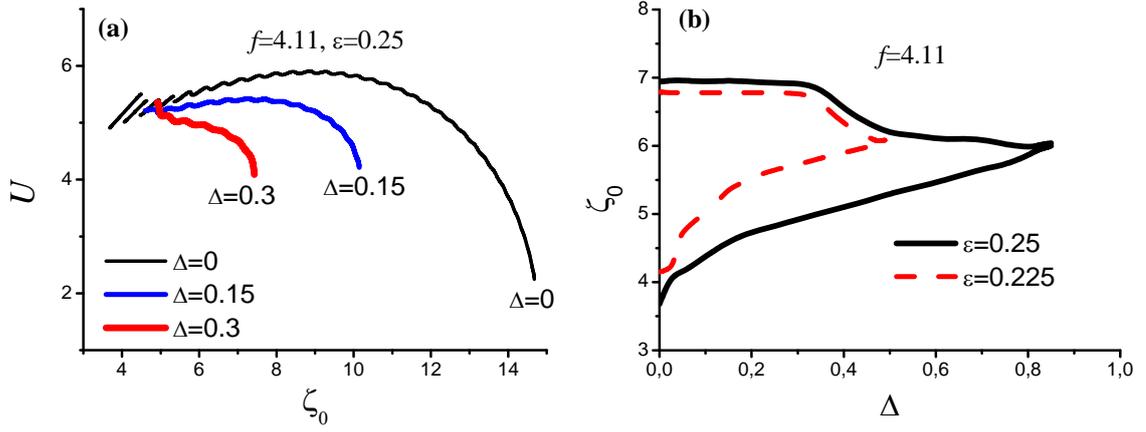

**Fig. 4.** (Color online) (a) Averaged intracavity intensity $U$ of platicons vs. normalized detuning $\zeta_0$ at $f = 4.11$, $\varepsilon = 0.25$ for different values of modulation frequency mismatch $\Delta$. (b) Excitation domain vs. mismatch value at $f = 4.11$, $\varepsilon = 0.25$ (black line) and $\varepsilon = 0.225$ (red line). $\varepsilon = 0.25$. In all cases $D_2 / \kappa \approx 1.036 \times 10^{-2}$.

Rewriting Eq. (1) in the following form

$$
\begin{aligned}
\frac{\partial a_\mu}{\partial \tau} = &-\left(1 + i\zeta_\mu\right)a_\mu + i \sum_{\mu' \leq \mu''} \left(2 - \delta_{\mu'\mu''}\right)a_{\mu'}a_{\mu''}a^{*}_{\mu'+\mu''-\mu} + \delta_{0\mu}f + \\
&+ \tilde{\varepsilon}\delta_{1\mu}f \exp\left(i\Delta\tau\right)
\end{aligned}
\tag{3}
$$

we also studied the usage of bi-harmonic pump for platicon generation. It was found that results obtained for bi-harmonic pump practically coincide with the results for amplitude-modulated pump if $\tilde{\varepsilon} \approx 2\varepsilon$. The main difference is that in contrast with amplitude-modulated pump biharmonic pump is sensitive to the sign of modulation frequency mismatch [compare Eqs. (1) and (3)]. Interestingly, platicon generation is possible even if there is a constant phase shift between two pumping waves.

To summarize, we showed that soft excitation (from noise-like inputs) of platicons is possible for amplitude-modulated or biharmonic pump if sideband pump power is large enough. For effective platicon excitation modulation frequency (or frequency difference between two pump waves) should be close to FSR of a microresonator. Excitation domain strongly depends on the sideband pump power and platicon duration may be altered significantly by tuning the pump frequency.

**REFERENCES**

[1] Del'Haye P., Schliesser A. *et al.*, *Nature*, **450** (2007) 1214.

[2] Savchenkov A. A., Matsko A. B. *et al.*, *Phys. Rev. Lett.*, **101** (2008) 93902.

[3] Kippenberg T. J., Holzwarth R. and Diddams S. A., *Science*, **332** (2011) 555.

[4] Li J., Lee H. *et al.*, *Phys. Rev. Lett.*, **109** (2012) 233901.

[5] Del'Haye P., Papp S. B., and Diddams S. A., *Phys. Rev. Lett.*, **109** (2012) 263901.

[6]   Pfeifle J., Brasch V. *et al.*, *Nat. Photon.*, **8** (2014) 375.

[7]   Ferdous F., Miao H. X., Leaird D. E., Srinivasan K., Wang J., Chen L., Varghese L. T., and Weiner A. M., *Nature Photon.*, **5** (2011) 770.

[8]   Del'Haye P., Herr T. *et al.*, *Phys. Rev. Lett.*, **107** (2011) 063901.

[9]   Herr T., Hartinger K. *et al.*, *Nature Photon.*, **6** (2012) 48.

[10] Herr T., and Brasch V. *et al.*, *Nature Photon.*, **8** (2014) 145.

[11] Brasch V., Herr T. *et al.*, http://arxiv.org/abs/1410.8598 (2014 ).

[12] Matsko A. B., Savchenkov A. A., and Maleki L., *Opt. Lett.*, **37** (2012) 43.

[13] Hansson T., Modotto D., and Wabnitz S., *Phys. Rev. A*, **88** (2013) 023819.

[14] Godey C., Balakireva I. *et al.*, *Phys. Rev. A*, **89** (2014) 063814.

[15] Coillet A., Balakireva I. *et al.*, *IEEE Photonics Journal*, **5**(4) (2013) 6100409.

[16] Liang W., Savchenkov A. A. *et al.*, *Opt. Lett.*, **39** (2014) 2920.

[17] Henriet R., Lin G. *et al.*, *Opt. Lett.* **40** (2015) 1567.

[18] Huang S. W., Zhou H. *et al.*, *Phys. Rev. Lett.*, **114** (2015) 053901.

[19] Xue X., Xuan Y. *et al.*, *Nat. Photon.* (2015).

[20] Lobanov V. E., Lihachev G. *et al.*, *Opt. Express*, **23** (2015) 7713.

[21] Xue X., Xuan Y. *et al.*, *Las. Photon. Rev.*, **9**  (2015) L-23.

[22] Herr T., Brasch V. *et al.*, *Phys. Rev. Lett.*, **113** (2014) 123901.

[23] Savchenkov A. A., Matsko A. B. *et al.*, *Opt. Express*, **20** (2012) 27290.

[24] Liu Y., Xuan Y. *et al.*, *Optica*, **1** (2014) 137.

[25] Strekalov D. V. and Yu N., *Phys. Rev. A*, **79** (2009) 041805(R).

[26] Hansson T. and Wabnitz S., *Phys. Rev. A*, **90** (2014) 013811.

[27] Antikainen A. and Agrawal G. P., *J. Opt. Soc. Am. B* **32** (2015) 1705.

[28] Chembo Y. K. K. and Yu N., *Phys. Rev. A*, **82** (2010) 33801.

[29] Hansson T., Modotto D., Wabnitz S., *Opt. Comm.*, **312** (2014) 134.

[30] Finot C., Kibler B. *et al.*, *J. Opt. Soc. Am. B*, **25** (2008)1938.

[31] Varlot B., Wabnitz S. *et al.*, *Opt. Lett.*, **38** (2013) 3899.

[32] Malaguti S., Bellanca G., and Trillo S., *Opt. Lett.,* **39** (2014) 2475.

[33] Taheri  H., Eftekhar A. A. *et al.*, *IEEE Phot. Journ.* **7** (2015) 2200309.

[34] Jang J. K., Erkintalo M. *et al.*, *Opt. Lett.* **40** (2015) 4755.

[35] Coen S. and Haelterman M., *Phys. Rev. Lett.*, **79** (1997) 4139.